\documentclass{aastex}
\usepackage{spr-astr-addons}
\RequirePackage{color}

\begin{document}
\title{Nano dust impacts on spacecraft and boom antenna charging\\
\textit{\small Accepted for publication in Astrophysics \& Space Science}}

\slugcomment{}
\shorttitle{Nano dust impacts on spacecraft}
\shortauthors{Pantellini et al}

\author{Filippo Pantellini} 
\and \author{Soraya Belheouane}
\and \author{Nicole Meyer-Vernet}
\and \author{Arnaud Zaslavsky}
\affil{1 LESIA, Observatoire de Paris, CNRS, UPMC,
Universit\'e Paris Diderot; 5 Place Jules Janssen, 92195
Meudon, France}


\begin{abstract}
High rate sampling detectors 
measuring the potential difference between the main body 
and boom antennas of interplanetary spacecraft have been shown to be
efficient means to measure the voltage pulses induced by nano dust impacts 
on the spacecraft body itself (see Meyer-Vernet et al, Solar Phys. {\bf 256}, 463 (2009)). 
However, rough estimates of the free charge  
liberated in post impact expanding plasma cloud indicate that 
the cloud's own internal electrostatic field is too weak to 
account for measured pulses as the ones from the TDS instrument  
on the STEREO spacecraft frequently exceeding $0.1\:{\rm V/m}$. 
In this paper we argue that the detected pulses 
are not a direct measure of the potential structure of the plasma cloud, 
but are rather the consequence 
of a transitional interruption of the photoelectron return current towards the
portion of the antenna located within the expanding cloud.  
\end{abstract}



\section{Introduction}

Dust particles in the nano and micro meter range hitting the main body of
interplanetary spacecraft have been shown to produce a transient potential 
difference between the antenna booms and the
spacecraft main body \citep[see][and references within]{Meyer-Vernet_Maksimovic_al_2009}. 
For large grains, in the micrometer range, 
the measured potential difference is primarily due to recollection of
the electrons from the impact generated expanding plasma cloud by the
spacecraft main body which under standard solar wind conditions at 1AU is
positively charged due to photoelectric
charging of its sunlight exposed parts. The temporary accumulation of a negative
charge $Q$ on the spacecraft's body accounts for a variation of the 
potential difference between the spacecraft and the antenna booms by an amount  
\begin{equation}
 \delta V=\Gamma\frac{Q}{C_{\rm SC}}\label{eq_dV}
\end{equation}
where $C_{\rm SC}$ is the spacecraft body capacitance and $\Gamma$ an order
unity gain factor. The potential pulse associated with
such large grain impacts being solely due to charging of the spacecraft body, 
one expects the signal $\delta V$ to be of similar amplitude on all antennas. 
Many thousands of such events displaying simultaneous and similar amplitude pulses
on different antennas have effectively been recorded on the three
antennas mounted on the STEREO spacecraft
\citep{Zaslavsky_Meyer-Vernet_al_2012}. 

Besides this group of events,    
STEREO has recorded an even larger sample of events where  
the voltage pulse is roughly equal on two antennas and  
larger, by a factor $\sim 20$, on the third one.
\cite{Zaslavsky_Meyer-Vernet_al_2012} interpret this second group of events 
as the signature of the impact of smaller and faster grains in the nano meter size
domain. 
As for the micro meter grains the two weak pulses are due to recollection of
electrons from the expanding plasma cloud and are an indirect measure of the
cloud's free charge content $Q$ via Equation (\ref{eq_dV}). 
As briefly described in Appendix A of \cite{Zaslavsky_Meyer-Vernet_al_2012}, 
the larger pulse measured on one of the three antennas is due to the
action of the expanding cloud's electric field on the photoelectrons surrounding
this particular antenna.  
In contrast, at most a small fraction of the two antennas presenting a weak signal 
is attained by the expanding cloud without a significant effect on the  
photoelectrons emitted by their surfaces. 
The reason this scenario does not apply in case of 
micro meter grain impacts is that the latter liberate a larger quantity of free charges than the
nano meter grain impacts so that the dissipation of the expanding cloud into the ambient plasma
occurs for cloud dimensions large enough ($\gtrsim 1m $) to envelop all three antennas.  

In this paper we present a semi-quantitative scenario to explain the strong 
positive charging of a boom antenna finding itself within a nano dust induced 
plasma cloud. We argue that the intrinsic electric field within the cloud is
too weak to account for the measured voltage pulse. However, this field
is shown to be strong enough to transitionally reduce the number of
photoelectrons falling back onto the antenna leading to a positive net current 
towards its surface. Except otherwise specified, SI units are used throughout
the paper.

\begin{figure}[h!]
\vspace*{2mm}
\begin{center}
\plotone{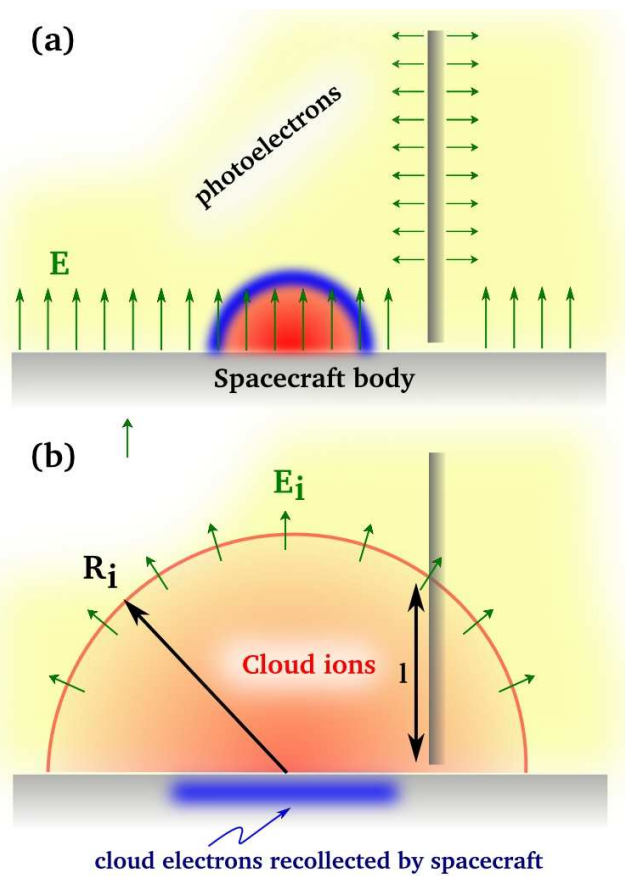}
\end{center}
\caption{A nanodust particle hitting a spacecraft's main body generates an
expanding plasma cloud. During the initial phase (a) the expanding cloud is
made of ions and electrons. Soon, for typical dimensions of the order of 
$10\:{\rm cm}$, the cloud's electrostatic field $E_{\rm i}$ becomes weaker than
the
spacecraft's own equilibrium field $E_{\rm SC}$. At that stage, cloud electrons 
become stripped to the spacecraft leaving a pure ion cloud continuing
expansion eventually encompassing a fraction of the antenna. 
We show that despite its weak intensity $E_{\rm i}\ll E_{\rm SC}$ the cloud's
electric field
prevents the photoelectron return current towards the antenna to compensate for
the emitted photoelectron current and the antenna increases its positive charge
beyond its equilibrium value.
}
\label{fig1}
\end{figure}

\section{Basic hypothesis and simplifications\label{sec_basic}}
We suppose that a dust grain hits the spacecraft body at some distance from
the base of an antenna boom in the idealised geometrical set-up  
illustrated in Figure \ref{fig1}a. 
For simplicity we assume that the spacecraft surface is plane with a cylindrical
antenna oriented perpendicularly to it. The spacecraft being exposed to the
solar radiation, both the spacecraft and the antenna are positively charged by
photoionisation with a permanent ``atmosphere'' of photoelectrons around them.
The characteristic energy of a photoelectron is of the order of 1 to 4 eV
\citep[see][and references within]{Henri_Meyer-Vernet_al_2011}
and the electric field $E$ near the sunlight exposed surfaces is a
few V/m. We emphasise that at 1 AU the photoelectron current from a sunlit surface 
of typical conducting spacecraft material 
is of the order $j_{\rm ph} = 5\: 10^{-5}{\rm A/m^2}$ per surface
unit normal to the Sun direction. This largely exceeds 
the flux of electrons provided by the solar wind which for a density $n_{\rm sw}$ of 10
electrons per ${\rm cm}^3$ and a temperature $T_{\rm e} =15{\rm eV}$ gives 
a 50 times smaller current (neglecting 
the effect of the spacecraft potential and the contribution from the ions)   
$j_{\rm sw} = e n_{\rm sw}(2\pi)^{-1/2}(k_{\rm B}T_{\rm e}/m)^{1/2} \approx 10^{-6}{\rm A/m^2}$, 
where $e$ is the elementary charge, $m$ the electron mass. 
Even considering that the solar wind electrons are collected at the same rate all
over the antenna surface whereas photoelectrons are only emitted by the one half of the antenna
exposed to solar radiation and that the  photoelectron flux is further reduced by a factor
$\sin\theta$ where $\theta$ is the angle of the antenna axis with respect to the direction of
the Sun, the emitted photoelectron current remains much stronger than the collected 
current from the ambient plasma. On STEREO the former is still 13 times stronger than the latter
\citep[see][]{Henri_Meyer-Vernet_al_2011}. 
To counter an endless charging of the spacecraft, most of the emitted photoelectrons (over $90\%$ 
for STEREO) are therefore doomed to fall back onto
the emitting surface. To simplify the discussion we 
then make the step to neglect the small proportion of photoelectrons 
escaping to infinity whose net current balances the currents provided by the ambient plasma. 
In any case, the zero photoelectron current assumption is not a crucial one 
as we do merely require that a substantial (not necessarily the totality) 
of the photoelectrons be recollected. 
 
The crucial point is that the extension of the cloud of
photoelectrons is much larger than the antenna radius  
$r_0\approx 1 {\rm cm}$ on STEREO. 
Indeed, a rough estimate of the Debye length 
of the photoelectron cloud can be obtained by assuming that the distribution of the 
radial velocities near the antenna surface is Maxwellian 
$f_{\rm ph}(v_r)=n_{\rm ph} \exp[-(v_r/v_{\rm ph})^2]/(v_{\rm ph}\sqrt{\pi})$ where 
$v_{\rm ph}=(2k_{\rm B}T_{\rm ph}/m)^{1/2}$ is the 
photoelectron thermal velocity. 
Equating the total current $2\pi r_0 L e\int_0^\infty v_r f_{\rm ph} dv_r$ directed away from the
antenna surface ($L$ is the length of the antenna) and the   
total photoelectron current from the antenna 
$j_{\rm ph} 2 r_0 L \sin\theta$ we obtain an estimate of the 
density $n_{\rm ph}$ near the antenna surface:
\begin{equation}
 n_{\rm ph}=\left(\frac{2}{\pi}\right)^{1/2}\:\left(\frac{m}{k_{\rm B}T_{\rm
ph}}\right)^{1/2}\:\frac{j_{\rm
ph}}{e} 
\: \sin\theta\label{eq_nph}
\end{equation}
Substituting the density 
$n_{\rm ph}$  into the expression for the Debye length $\lambda_{\rm ph}^2 = \varepsilon_0 k_{\rm B}
T_{\rm ph}/(e^2 n_{\rm ph})$ yields  
\begin{equation}
 \lambda_{\rm ph}^2 \approx 
\frac{\varepsilon_0 m}{e j_{\rm ph} \sin\theta}
\left(\frac{\pi}{2}\right)^{1/2} 
\left(\frac{k_{\rm B}T_{\rm ph}}{m}\right)^{3/2} \label{eq_lph}
\end{equation}
where $\varepsilon_0$ is the vacuum permittivity and  $k_{\rm B}$ 
the Boltzmann constant.
Setting $T_{\rm ph}=2{\rm eV}$ and taking an average value $2/\pi$ 
for $|\sin\theta|$ one has $\lambda_{\rm ph}=0.64 {\rm m}$ which is indeed 
much larger than the antenna radius $r_0$. We note in passing that the 
Debye length of the ambient plasma at 1AU is typically $\gtrsim 10{\rm m} \gg \lambda_{\rm ph}$ 
so that it can be safely ignored in the present context.

After a dust impact on the spacecraft body, as schematically illustrated in Figure 
\ref{fig1}a, a hemispherical overall neutral plasma cloud made of ions (red) and
electrons (blue) and neutrals expands away from the impact point. In the early
phase of the expansion the electric field intensities within the cloud are
stronger than the $\sim$V/m field intensity surrounding the spacecraft and the
expansion is not affected by the environment. At some stage the electric field within the
expanding cloud has decreased below the spacecraft's field. The cloud
electrons become captured by the spacecraft leaving a positively charged cloud
continue the expansion 
eventually encompassing a non negligible portion of the antenna
(Figure \ref{fig1}b). In the next section we evaluate the field intensity
$E_{\rm i}$ within the cloud and argue that this field is strong enough to
temporally prevent a significant fraction of photoelectrons to fall 
back onto the antenna.

\subsection{Early phase of the expansion}

Let us assume that a dust grain of $10^{-20}{\rm kg}$ hits the spacecraft body 
at a velocity of 400 km/s.
The empirical formula (2) in \cite{McBride_McDonnell_1999} 
predicts that the post impact released free charge is $Q\approx 3\:
10^{-12}{\rm C}$. For comparable ion and electron temperatures, electrons tend 
to detach from the ions forming an electron precursor as illustrated in Figure
\ref{fig1}a \citep[e.g.][]{Pantellini_Landi_al_2012}. 
The strongest possible electric field intensity
sensed by the
electrons in the precursor is obtained in the limiting case 
of complete electron-ion charge separation. If $R_{\rm i}$ is the 
radius of the ion sphere, the maximum field intensity near its edge 
is at most $E_{\rm i,max} = Q/4\pi\epsilon_0 R_{\rm i}^2$, or less in case 
of partial charge separation. Even in the limit of complete charge
separation this field is smaller than the typical spacecraft electric field
intensity of ~5V/m by the time the cloud has grown to 
a small radius of $R_{\rm i}=7{\rm cm}$ only. 
Upon further expansion its intrinsic field falls below the spacecraft's own field 
which then starts recollecting electrons from within the cloud.  
The time $t_0$ for an electron leaving the spacecraft with a normal velocity $v_0$ 
to reach its maximum height $h_{\rm max}$ can be estimated by assuming that it only feels  
the constant spacecraft electric field, i.e. $t_0 \sim v_0 m/eE_{\rm SC}$. The non constant cloud
field is also directed towards the
spacecraft and may further reduce $t_0$. 
For an electron with an initial energy of the order 
the photoelectron thermal energy $3k_{\rm B}T_{\rm ph}/2$ 
and under the assumption of energy equipartition we have $v_0^2=k_{\rm B}T_{\rm ph}/m$. 
Setting $E_{\rm SC}=5{\rm V/m}$  and, as before, $T_{\rm ph}=2{\rm eV}$ 
it then takes a time $t_0\sim 0.67 \mu{\rm s}$ to reach a maximum height  
above the spacecraft $h_{\rm max}\sim v_0 t_0/2=0.2{\rm m}$.
This is an upper estimate for $h_{\rm max}$ as we have assumed that the electrons are
collisionless and insensitive to the cloud's field. In addition,   
cloud electrons are expected to cool during
expansion \citep[e.g.][]{Murakami_Basko_2006,Beck_Pantellini_2009},    
which further favours fast recollection. 
The bottom line is that the electrons of the
cloud are recollected by the spacecraft before its maximum extension 
$R_{\rm i,max}$ has been reached, i.e. before its density has
decreased to a value comparable to the surrounding solar wind plasma density. 
Indeed, for a spherical cloud of charge $Q=3\:10^{-12}{\rm C}$, and a solar wind
density $n_{\rm SW}\approx 5 {\rm cm^{-3}}$ the relation $Q/e = n_{\rm SW}
(4\pi/3)
R_{\rm i,max}^3$ gives $R_{\rm i,max}\approx 1 {\rm m}$. 
The late evolution of the
cloud, when the chance of having a significant portion of the antenna within
the cloud itself is high, can be assumed to be hemispherically shaped
as shown in Figure \ref{fig1}b. At this stage, all electrons from the cloud 
have been recollected by the spacecraft leaving 
the cloud with a total positive charge $Q$.

The electric field $E_{\rm i}$ near the edge of a spherical cloud 
$R_{\rm i}$ is then given by the Coulomb potential 
\begin{equation}
 E_{\rm i} = \frac{Q}{4\pi\epsilon_0 R_{\rm i}^2}\label{eq_Ei}.
\end{equation}
We emphasise that the field $E_{\rm i}$ given by (\ref{eq_Ei})  
is the cloud's field after its 
electrons having been recollected by the spacecraft, which is by definition   
smaller than the spacecraft field $E_{\rm SC}$. 
Because of the $E_{\rm i} \propto R_{\rm i}^{-2}$ dependence 
one can even assume that during this late phase of the expansion $E_{\rm i}\ll
E_{\rm SC}$. We shall see that despite being small, the cloud's field $E_{\rm
i}$ is generally strong enough to drastically reduce the photoelectron return
current towards the antenna inducing a transitional modification of the
antenna's net charge.

\subsection{Photoelectron dynamics}
 
During the time periods between successive impacts we may assume a time
independent electrostatic potential $U$ in the plasma surrounding 
the antenna. This assumption holds for plasma conditions that vary slower than
both the inverse of the ambient plasma frequency $\omega_{\rm e}^{-1}$ 
and the photoelectron plasma frequency $\omega_{\rm ph}^{-1}$ . 
The total energy of a photoelectron in such a static field 
can be written as 
\begin{equation}
 \varepsilon = \frac{1}{2}\: m (v_r^2+v_z^2) + \frac{M^2}{2mr^2} +eU(r)={\rm
constant}\label{eq_E}
\end{equation}
where $r$ is the distance to the 
antenna axis, $v_r$ the radial component of the velocity, $v_z$  the (constant)
velocity component along the antenna axis and
$M=mrv_\theta$ the angular momentum with $v_\theta$ being the azimuthal velocity
component. Both the total
energy $\varepsilon$ and the angular momentum $M$ are conserved quantities as
long as $U$ is time independent. 
As already pointed out, a 
majority of the photoelectrons emitted by the antenna must fall back onto its  
surface as this is the only way to balance the net (outflowing + inflowing) photoelectron current 
from the antenna and the currents from the ambient plasma. 
The size of the cloud formed by these ballistic photoelectrons is expected to be much larger than
the antenna radius $r_0$ (cf Figure \ref{fig1}), since the photoelectron Debye length estimated in
(\ref{eq_lph}) is $\gg r_0 $. 

\begin{figure}[h!]
\vspace*{2mm}
\begin{center}
\plotone{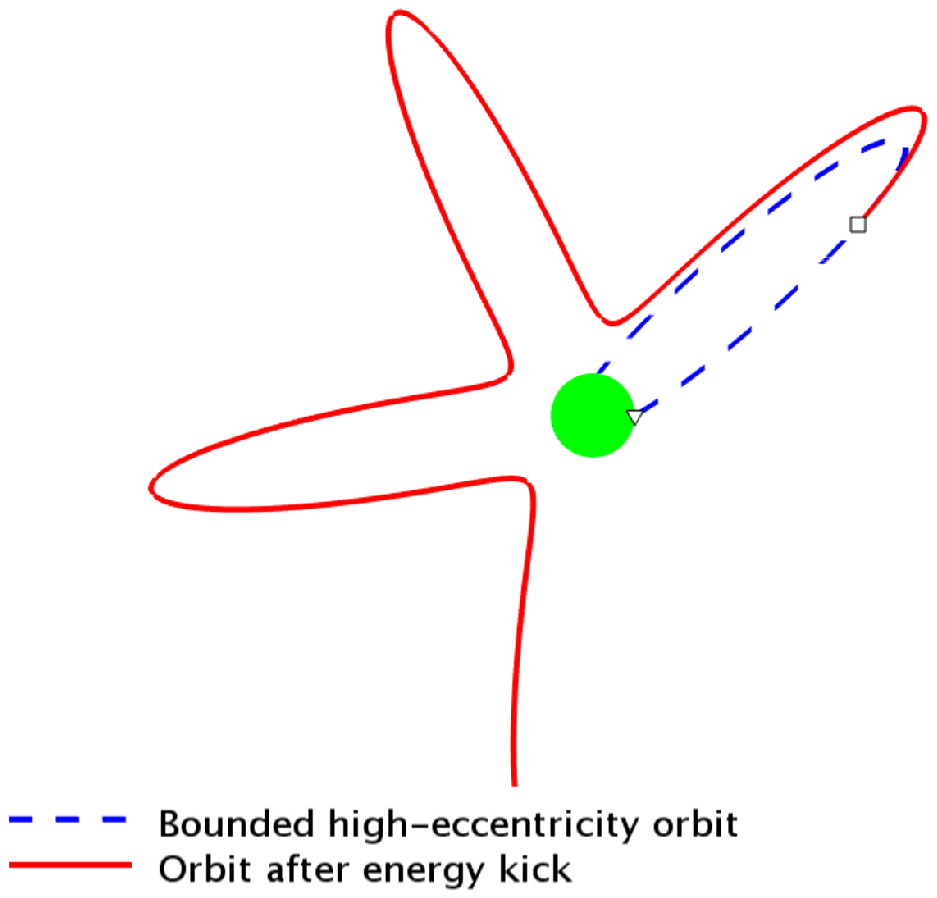}
\end{center}
\caption{Top view showing the section of a boom antenna and a typical
photoelectron orbit. Under quite conditions the electrostatic field around the
positively charged antenna is axisymmetric and a bounded photoelectron emitted
at its surface (triangle) comes back to the antenna along the dashed (blue)
line. Given the high eccentricity of the orbit, the electron spends most of its
time at large distances from the antenna and its velocity transverse to the
radial direction is necessarily small. Thus even a minor energy kick (at the
place
marked by a square) prevents the electron from falling back onto the antenna. 
We argue that the electrostatic field from a   
nano dust impact generated expanding plasma cloud is  
a sufficiently strong perturbation 
to prevent a significant fraction of photoelectrons 
to return to the surface, forcing
a rapid and measurable charging of the antenna.
}
\label{fig2}
\end{figure}
The only way for photoelectrons to be emitted and recollected by 
a thin antenna while forming an extended cloud much larger than $r_0$ 
is to move on high eccentricity 
orbits similar to the blue (dashed) curve in figure \ref{fig2}.
As $r\gg r_0$ on most of a typical photoelectron trajectory, 
$M^2/2mr^2$ is negligible with respect to $M^2/2mr_0^2$ and 
equation (\ref{eq_E}) can be approximated by
an equation for the radial component of the velocity only:   
\begin{equation}
  \frac{1}{2}
m v_r^2+e[U(r)-U(r_0)]\approx 
\frac{1}{2}m(v_{r0}^2+v_{\theta0}^2),\;\;\;{\rm for}\;r\gg
r_0\label{E_approx}
\end{equation}
where we have assumed that there is no electric field along the $z$-axis so that 
$v_z={\rm constant}$. 
Averaging (\ref{E_approx}) over a large number of trajectories, and setting
$U(r_0)=0$, leads to the formal expression $\langle m v_r ^2 /2\rangle +\langle
e U(r)\rangle=\langle m(v_{r0}^2 +v_{\theta0}^2)/2\rangle$.
Explicit computation of averages
requires a detailed knowledge of the distribution of the electrons injected at
$r_0$ and the potential energy profile $U(r)$. We limit ourselves to an order
of magnitude estimate by assuming equipartition between potential 
and kinetic energy $\langle K_r\rangle\equiv\langle m v_r ^2 /2\rangle \approx
\langle eU(r)\rangle$ and an energy of $k_{\rm B}T_{\rm ph}/2$ per degree of freedom at $r=r_0$, 
i.e. $\langle K_{r,0}\rangle\approx k_{\rm B}T_{\rm ph}/2 $.

On the other hand, conservation of the angular momentum conservation $M$ implies 
that the azimuthal kinetic energy $K_{\theta}=m v_\theta ^2 /2$ decreases 
with distance as $(r_0/r)^2$.  Assuming the same characteristics for the 
photoelectron cloud as in Section \ref{sec_basic} and 
given an antenna radius $r_0\approx 1{\rm cm}$, an average kinetic
energy at $r_0$ of $K_{\theta,0}=k_{\rm B}T_{\rm ph}/2 =1{\rm eV}$ and an mean position
$\langle r\rangle=\lambda_{\rm ph}/2=0.32 {\rm m}$ one has $K_\theta(0.32{\rm m})\approx 
10^{-3}{\rm eV}$. 
The average azimuthal energy of the photoelectron is
so small that it can be easily increased by a factor larger than unity 
in the field of the expanding plasma cloud. Let us verify this statement.  
As previously noted a nano dust of $10^{-20}$kg hitting a spacecraft  
at $400$km/s  produces a cloud with a free 
charge $Q=3\:10^{-12}{\rm C}$. In order for the 
electrostatic voltage $Q/4\pi\epsilon_0 R_{\rm i}$ to be equal to
$K_\theta/e$ requires $R_{\rm i}\approx
30 {\rm m}$ which is well beyond the size of the cloud at the time 
it merges with the surrounding plasma. We note that even if the liberated free 
charge $Q$ was only one tenth of the above estimate 
based on the empirical formula (9) in \cite{Meyer-Vernet_Maksimovic_al_2009},
the intrinsic field of the cloud would still be strong enough to 
disconnect most of the photoelectrons from the antenna surface.  
In other words, during the whole time of the expansion, the cloud's electrostatic field   
is potentially strong enough  
to increase the azimuthal velocity of the ballistic photoelectrons by a
factor much larger than unity, while  letting their total kinetic energy essentially unaffected. 
Only in the very special case of a cloud centred exactly at the base of the antenna 
the azimuthal field component is zero and the mechanism inoperative. 

A qualitative illustration of the effect of a slight modification of the 
total kinetic energy of a photoelectron on a high-eccentricity orbit is shown
in Figure \ref{fig2}. 
A photoelectron is emitted at the surface of the
positively charged antenna of radius $r_0$ with enough energy to escape to
a distance $r/r_0\gg 1$. If the electrostatic potential of the antenna is time
independent and if the total energy of the electron is negative, the 
electron is doomed to fall back onto the antenna's surface (blue dashed line). 
Its contribution to the net current from the antenna is therefore zero. However,
a small energy kick to the electron at an arbitrary position 
(marked by a square on the figure) may be sufficient to alter the orbit such that it does
no longer cross the antenna's surface. The emitted electron being unable to return towards the
antenna, the latter increases its charge by a positive elementary charge $e$ 
corresponding to a net positive current. 

In Figure \ref{fig2} the electrostatic field is given by the
modified Bessel function $U(r) \propto K_0(r/\mathcal{R})$ where
$\mathcal{R}$ can then be seen as the characteristic radial extension of the 
photoelectron cloud. 
Such a field may represent a fair approximation of the electrostatic field
around a positively charged antenna in the linear regime $eU/k_{\rm B}T_{\rm ph}\ll 1$
\citep{Bystrenko_Bystrenko_2008}, i.e. 
at some distance from the antenna but is certainly incorrect near $r_0$
where potential and kinetic energy must be of same order. 
Without pretending to reproduce in detail a real case, which is not  
the purpose of the present work, we have arbitrarily set  
$r_0/\mathcal{R}=0.1$ and the electron has been injected 
at the antenna surface at a $30^o$ angle with respect to the radial direction
and an energy corresponding to 0.83 times the escape energy. An energy  
kick in the transverse direction corresponding to 0.05 times the initial kinetic
energy is given to the electron at the point (marked by a square) where its
radial velocity has decreased to ~1/3 of its initial value.  

\subsection{Antenna charging}

Let us suppose, as illustrated in Figure \ref{fig1}b that $l$ represents the
length of the part of the antenna located inside the expanding plasma cloud at
the time of its maximum extension, before it merges with the ambient plasma.
Assuming that the orbits of the majority of the photoelectrons emitted by the
sunlit surface of the portion of antenna located inside the cloud become
orbitally disconnected from the antenna's surface during a time $\tau$, 
the excess charge $Q$ which accumulates on the antenna can be estimated to  
\begin{equation}
 Q_{\rm a}(l) \sim j_{\rm ph} 2 r_0 l \:\tau\label{eq_Qa}.
\end{equation}
In (\ref{eq_Qa}) $j_{\rm ph}=5\:10^ {-5} {\rm A/m^2}$ is a typical 
photoelectron current at 1AU and $\tau$ is a characteristic time for the
population of photoelectrons to restore 
the original axisymmetric antenna potential. 
The smallest possible value for $\tau$ is determined by the 
inverse of the photoelectron plasma frequency $\omega_{\rm
ph}$, which is the fastest collective time scale for the photoelectrons, 
so that $\tau \sim 2\pi /\omega_{\rm
ph}\propto n_{\rm ph}^{-1/2}$. The 
time interval $\tau$, during which the antenna increases its positive 
charge, is a function of the spatially varying photoelectron density 
$n_{\rm ph}$. The spatial dependence of the photoelectron density  
makes it difficult to give a number to feed into (\ref{eq_Qa}). Assuming 
an average density $n_{\rm ph}=100 {\rm cm}^{-3}$ for the extended 
cloud (to be compared with the estimate at the surface $n_{\rm ph}(r_0)=267 {\rm cm}^{-3}$ 
from equation (\ref{eq_nph})) 
one has $\tau\sim 11 \mu{\rm s}$ and an
estimate $Q_{\rm a}/l\approx 1.7\: 10^{-11} {\rm C/m}$. 
The induced voltage pulse $\delta V$ for an antenna of 
capacitance $C_{\rm a}$ and a gain $\Gamma$ is given by 
\begin{equation}
 \delta V = \Gamma\:\frac{Q_{\rm a}(l)}{C_{\rm a}}.\label{eq_dVa}
\end{equation}
For the 6m antennas mounted on STEREO $C_{\rm a}\approx 63{\rm pF}$ and
$\Gamma\approx 0.5$ \citep{Bale_Ullrich_al_2008} the expected voltage
pulse may then be as large as $\delta V_{\rm
ST}\approx 0.3\: l$ with $[l]={\rm m}$ and $[V_{\rm ST}]={\rm V}$. Taking $l=0.5{\rm
m}$ for a cloud of radius $R_{\rm i,max}\approx 1
{\rm m}$, the expected voltage pulse is $\delta V_{\rm
ST}=70{\rm mV}$ which is of the order of the average pulses measured 
on STEREO for the so-called single hits 
\citep[see figure A.10 in][]{Zaslavsky_Meyer-Vernet_al_2012}. 

From equation (\ref{eq_Qa}) it appears that the charge collected by the 
antenna is proportional to its radius $r_0$. On the WIND spacecraft, also  
located at 1AU from the Sun $r_0$ is approximately  
60 times smaller than on STEREO \citep{Kellogg_Bale_2001} which implies that the voltage pulses 
associated with impacts of nano dusts are expected to be smaller by this same 
factor with a typical expected amplitude of $1{\rm mV}$, only. 
Such pulses are too weak to emerge from the natural electrostatic plasma fluctuations 
and are therefore undetectable on WIND, not even considering that 
WIND's antennas have a larger capacitance than STEREO's.  
Dust impact detections on the Cassini spacecraft at Saturn, which carries radio instruments
similar to STEREO are also problematic as the photoelectron current 
at Saturn is roughly 80 times smaller than at 1 AU with a commensurately smaller signal. 

\section{Conclusions}

Estimates of the electrostatic potential through  
a plasma cloud generated by nano meter sized dust grain impacts on 
a spacecraft's main body suggest that the cloud's field is 
too weak to account for the voltage pulses observed on STEREO 
\citep[see][]{Pantellini_Landi_al_2012}.
To solve the issue we suggest that the strong voltage pulses measured between one 
individual boom antenna and STEREO's main body is primarily the consequence of
a charging of the antenna due to a temporary interruption of the photoelectron
return current. The interruption of the return current only affects the
fraction of the antenna finding itself within the plasma cloud at the time of
its maximum expansion. This happens because the 
photoelectron ``atmosphere'' bounded to the antenna extends to distances much
larger than the antenna radius $r_0$ itself. This is characteristic 
of situations where the photoelectron Debye length is $\gg r_0$.
Under such circumstances most photoelectrons emitted by the antenna have high
eccentricity orbits, meaning that on most of their trajectory their velocity is
essentially oriented radially with respect to the antenna axis. Angular moment 
conservation implies that the azimuthal velocity component of a photoelectron
decreases with distance as $r_0/r$ and, consequently, its azimuthal energy as
$(r_0/r)^2$. As most photoelectrons 
are located at a large distance from the antenna $r\gg r_0$, even the small
energy kick given to them by the expanding plasma cloud can be strong enough to
change their azimuthal velocity (and therefore their angular momentum) by a
factor larger than unity. Such a strong increase of the angular momentum is
generally sufficient to disconnect the corresponding photoelectron from 
its ballistic trajectory connected to the
antenna's surface which therefore undergoes a net loss of negative charges. 

The positive charging of the antenna  
continues until both the perturbing cloud becomes diluted in the
ambient plasma and the cylindrical symmetry of the potential around the
antenna is reestablished, i.e. at least during a time of the order of a photoelectron
plasma oscillation. 

This scenario is compatible with the fact that nano dust impacts 
are not readily detectable on radio spectrograms from radio receivers on   
WIND at Earth orbit and Cassini at Saturn,  
as in both cases the emitted photoelectron current 
is strongly reduced due to either a smaller antenna radius (WIND) or a
larger distance from the Sun (Cassini) than STEREO.   


\begin{thebibliography}{10}
\ifx \bisbn   \undefined \def \bisbn  #1{ISBN #1}\fi
\ifx \binits  \undefined \def \binits#1{#1} \fi
\ifx \bauthor  \undefined \def \bauthor#1{#1} \fi
\ifx \batitle  \undefined \def \batitle#1{#1} \fi
\ifx \bjtitle  \undefined \def \bjtitle#1{#1}\fi
\ifx \bvolume  \undefined \def \bvolume#1{\textbf{#1}}\fi
\ifx \byear  \undefined \def \byear#1{#1} \fi
\ifx \bissue  \undefined \def \bissue#1{#1} \fi
\ifx \bfpage  \undefined \def \bfpage#1{#1} \fi
\ifx \blpage  \undefined \def \blpage #1{#1} \fi
\ifx \burl  \undefined \def \burl#1{\textsf{#1}} \fi
\ifx \doiurl  \undefined \def \doiurl#1{\textsf{#1}} \fi
\ifx \betal  \undefined \def \betal{\textit{et al.}} \fi
\ifx \binstitute  \undefined \def \binstitute#1{#1} \fi
\ifx \binstitutionaled  \undefined \def \binstitutionaled#1{#1} \fi
\ifx \bctitle  \undefined \def \bctitle#1{#1} \fi
\ifx \beditor  \undefined \def \beditor#1{#1} \fi
\ifx \bpublisher  \undefined \def \bpublisher#1{#1} \fi
\ifx \bbtitle  \undefined \def \bbtitle#1{#1} \fi
\ifx \bedition  \undefined \def \bedition#1{#1} \fi
\ifx \bseriesno  \undefined \def \bseriesno#1{#1} \fi
\ifx \blocation  \undefined \def \blocation#1{#1} \fi
\ifx \bsertitle  \undefined \def \bsertitle#1{#1} \fi
\ifx \bsnm \undefined \def \bsnm#1{#1} \fi
\ifx \bsuffix \undefined \def \bsuffix#1{#1} \fi
\ifx \bparticle \undefined \def \bparticle#1{#1} \fi
\ifx \barticle \undefined \def \barticle#1{#1} \fi
\ifx \bconfdate \undefined \def \bconfdate #1{#1} \fi
\ifx \botherref \undefined \def \botherref #1{#1} \fi
\ifx \url \undefined \def \url#1{\textsf{#1}} \fi
\ifx \bchapter \undefined \def \bchapter#1{#1} \fi
\ifx \bbook \undefined \def \bbook#1{#1} \fi
\ifx \bcomment \undefined \def \bcomment#1{#1} \fi
\ifx \oauthor \undefined \def \oauthor#1{#1} \fi
\ifx \citeauthoryear \undefined \def \citeauthoryear#1{#1} \fi
\ifx \endbibitem  \undefined \def \endbibitem {}\fi
\ifx \bconflocation  \undefined \def \bconflocation#1{#1} \fi
\ifx \arxivurl  \undefined \def \arxivurl#1{\textsf{#1}} \fi

\bibitem[\protect\citeauthoryear{Bale et~al.}{2008}]{Bale_Ullrich_al_2008}
\begin{barticle}
\bauthor{\bsnm{Bale}, \binits{S.D.}},
\bauthor{\bsnm{Ullrich}, \binits{R.}},
\bauthor{\bsnm{Goetz}, \binits{K.}},
\bauthor{\bsnm{Alster}, \binits{N.}},
\bauthor{\bsnm{Cecconi}, \binits{B.}},
\bauthor{\bsnm{Dekkali}, \binits{M.}},
\bauthor{\bsnm{Lingner}, \binits{N.R.}},
\bauthor{\bsnm{Macher}, \binits{W.}},
\bauthor{\bsnm{Manning}, \binits{R.E.}},
\bauthor{\bsnm{{McCauley}}, \binits{J.}},
\bauthor{\bsnm{Monson}, \binits{S.J.}},
\bauthor{\bsnm{Oswald}, \binits{T.H.}},
\bauthor{\bsnm{Pulupa}, \binits{M.}}:
\bjtitle{Space Science Reviews}
\bvolume{136},
\bfpage{529}
(\byear{2008})
\end{barticle}
\endbibitem

\bibitem[\protect\citeauthoryear{Beck and
  Pantellini}{2009}]{Beck_Pantellini_2009}
\begin{barticle}
\bauthor{\bsnm{Beck}, \binits{A.}},
\bauthor{\bsnm{Pantellini}, \binits{F.}}:
\bjtitle{Plasma Physics and Controlled Fusion}
\bvolume{51}(\bissue{1}),
\bfpage{015004}
(\byear{2009})
\end{barticle}
\endbibitem

\bibitem[\protect\citeauthoryear{Bystrenko and
  Bystrenko}{2008}]{Bystrenko_Bystrenko_2008}
\begin{barticle}
\bauthor{\bsnm{Bystrenko}, \binits{O.}},
\bauthor{\bsnm{Bystrenko}, \binits{T.}}:
\bjtitle{Physica Scripta}
\bvolume{78}(\bissue{2}),
\bfpage{025502}
(\byear{2008})
\end{barticle}
\endbibitem

\bibitem[\protect\citeauthoryear{Henri
  et~al.}{2011}]{Henri_Meyer-Vernet_al_2011}
\begin{barticle}
\bauthor{\bsnm{Henri}, \binits{P.}},
\bauthor{\bsnm{{Meyer-Vernet}}, \binits{N.}},
\bauthor{\bsnm{Briand}, \binits{C.}},
\bauthor{\bsnm{Donato}, \binits{S.}}:
\bjtitle{Physics of Plasmas}
\bvolume{18},
\bfpage{2308}
(\byear{2011})
\end{barticle}
\endbibitem

\bibitem[\protect\citeauthoryear{Kellogg and Bale}{2001}]{Kellogg_Bale_2001}
\begin{barticle}
\bauthor{\bsnm{Kellogg}, \binits{P.J.}},
\bauthor{\bsnm{Bale}, \binits{S.D.}}:
\bjtitle{Journal of Geophysical Research}
\bvolume{106}(\bissue{A9}),
\bfpage{18721}
(\byear{2001})
\end{barticle}
\endbibitem

\bibitem[\protect\citeauthoryear{{McBride} and
  {McDonnell}}{1999}]{McBride_McDonnell_1999}
\begin{barticle}
\bauthor{\bsnm{{McBride}}, \binits{N.}},
\bauthor{\bsnm{{McDonnell}}, \binits{J.A.M.}}:
\bjtitle{Planetary and Space Science}
\bvolume{47},
\bfpage{1005}
(\byear{1999})
\end{barticle}
\endbibitem

\bibitem[\protect\citeauthoryear{{Meyer-Vernet}
  et~al.}{2009}]{Meyer-Vernet_Maksimovic_al_2009}
\begin{barticle}
\bauthor{\bsnm{{Meyer-Vernet}}, \binits{N.}},
\bauthor{\bsnm{Maksimovic}, \binits{M.}},
\bauthor{\bsnm{Czechowski}, \binits{A.}},
\bauthor{\bsnm{Mann}, \binits{I.}},
\bauthor{\bsnm{Zouganelis}, \binits{I.}},
\bauthor{\bsnm{Goetz}, \binits{K.}},
\bauthor{\bsnm{Kaiser}, \binits{M.L.}},
\bauthor{\bsnm{Cyr}, \binits{O.C.S.}},
\bauthor{\bsnm{Bougeret}, \binits{J.}},
\bauthor{\bsnm{Bale}, \binits{S.D.}}:
\bjtitle{Solar Physics}
\bvolume{256},
\bfpage{463}
(\byear{2009})
\end{barticle}
\endbibitem

\bibitem[\protect\citeauthoryear{Murakami and
  Basko}{2006}]{Murakami_Basko_2006}
\begin{barticle}
\bauthor{\bsnm{Murakami}, \binits{M.}},
\bauthor{\bsnm{Basko}, \binits{M.M.}}:
\bjtitle{Physics of Plasmas}
\bvolume{13},
\bfpage{012105}
(\byear{2006})
\end{barticle}
\endbibitem

\bibitem[\protect\citeauthoryear{{Pantellini}
  et~al.}{2012}]{Pantellini_Landi_al_2012}
\begin{barticle}
\bauthor{\bsnm{{Pantellini}}, \binits{F.}},
\bauthor{\bsnm{{Landi}}, \binits{S.}},
\bauthor{\bsnm{{Zaslavsky}}, \binits{A.}},
\bauthor{\bsnm{{Meyer-Vernet}}, \binits{N.}}:
\bjtitle{Plasma Physics and Controlled Fusion}
\bvolume{54}(\bissue{4}),
\bfpage{045005}
(\byear{2012}).
doi:\doiurl{10.1088/0741-3335/54/4/045005}
\end{barticle}
\endbibitem

\bibitem[\protect\citeauthoryear{Zaslavsky
  et~al.}{2012}]{Zaslavsky_Meyer-Vernet_al_2012}
\begin{barticle}
\bauthor{\bsnm{Zaslavsky}, \binits{A.}},
\bauthor{\bsnm{{Meyer-Vernet}}, \binits{N.}},
\bauthor{\bsnm{{Mann}}, \binits{I.}},
\bauthor{\bsnm{{Czechowski}}, \binits{A.}},
\bauthor{\bsnm{Issautier}, \binits{K.}},
\bauthor{\bsnm{{Le Chat}}, \binits{G.}},
\bauthor{\bsnm{Pantellini}, \binits{F.}},
\bauthor{\bsnm{Goetz}, \binits{K.}},
\bauthor{\bsnm{Maksimovic}, \binits{M.}},
\bauthor{\bsnm{Bale}, \binits{S.D.}},
\bauthor{\bsnm{Kasper}, \binits{J.C.}}:
\bjtitle{Journal of Geophysical Research}
\bvolume{in press}
(\byear{2012}).
doi:\doiurl{10.1029/2011JA017480}
\end{barticle}
\endbibitem

\end{thebibliography}

\end{document}